\documentclass[12pt, a4paper]{elsarticle}
\usepackage[T1]{fontenc}
\usepackage[utf8]{inputenc}
\usepackage[italian,english]{babel}
\usepackage{amsmath, amsfonts, amssymb}
\usepackage{float}
\usepackage[font=footnotesize]{caption}
\usepackage{graphicx}
\usepackage{epstopdf}
\usepackage{upgreek}
\usepackage{geometry}
\usepackage{subfig}
\usepackage{caption}
\usepackage{textcomp}
\usepackage{color}

\renewcommand{\vec}[1]{\boldsymbol{#1}}

\def \curl{\mbox{curl\hskip 1pt}}
\def \Curl{\mbox{Curl\hskip 1pt}}
\def \div{\mbox{div\hskip 1pt}}
\def \Div{\mbox{Div\hskip 1pt}}
\def \tr{\mbox{tr\hskip 1pt}}
\def \grad{\mbox{grad\hskip 1pt}}
\def \Grad{\mbox{Grad\hskip 1pt}}

\journal{International Journal of Engineering Science}

\title{Stability analysis of charge-controlled \\ soft dielectric plates}

\author{Hannah Conroy Broderick$^{1}$, Michele Righi$^2$, Michel Destrade$^1$, \\
Ray W. Ogden$^3$\\[12pt]
$^1$ \normalsize{School of Mathematics, Statistics and Applied Mathematics,} \\
\normalsize{NUI Galway, University Road, Galway, Ireland}\\[8pt]
$^2$ \normalsize{TeCip Istituto, Scuola Superiore Sant'Anna,} \\
\normalsize{Piazza Martiri della Libertà 33, 5612, Pisa, Italy}\\[8pt]
$^3$ \normalsize{School of Mathematics and Statistics,} \\
\normalsize{University of Glasgow, University Place, Glasgow G12 8SQ, Scotland, UK}\\[12pt]
 }

\date{March 2020}

\begin{document}

\begin{frontmatter}

\begin{abstract}
We examine the stability of a soft dielectric plate deformed by the coupled effects of a mechanical pre-stress applied on its lateral faces and an electric field applied through its thickness under charge control.
The electric field is created by spraying charges on the major faces of the plate: although in practice this mode of actuation is harder to achieve than a voltage-driven deformation, here we find that  it turns out to be much more stable in theory and in simulations. 

First we show that the electromechanical instability based on the Hessian criterion associated with the free energy of the system does not occur at all for charge-driven dielectrics for which the electric displacement is linear in the electric field.
Then we show that the geometric instability associated with the formation of small-amplitude wrinkles on the faces of the plate that arises under voltage control does not occur either under charge control. This is in complete contrast to voltage-control actuation, where Hessian and wrinkling instabilities can occur once certain critical voltages are reached. 

For the mechanical pre-stresses, two modes that can be implemented in practice are used: equi-biaxial and uni-axial.
We confirm the analytical  and numerical stability results of homogeneous deformation modes with Finite Element simulations of  real actuations, where inhomogeneous fields may develop. 
We find complete agreement in the equi-biaxial case, and very close agreement in the uni-axial case, when the pre-stress is due to a dead-load weight. In the latter case, the simulations show that small inhomogeneous effects develop near the clamps, and eventually a compressive lateral stress emerges, leading to a breakdown of the numerics.

\end{abstract}


\end{frontmatter}


\section{Introduction}


Soft dielectric materials can undergo large actuation stretches when a potential difference is induced in the material.
Typically, compliant electrodes such as carbon grease are smeared onto the faces of a soft dielectric elastomer plate and a voltage is applied across the thickness of the material.  As the voltage increases the material gradually expands in area until a maximum voltage is reached, at which point a rapid large deformation known as \emph{snap-through} occurs
\cite{ZhSu07}.
The large actuation achieved due to the snap-through behaviour is desirable for many applications but is difficult to achieve in practice.
Snap-through is often prevented by electric breakdown \cite{ZhSu07} or by instabilities such as  inhomogeneities \cite{Bertoldi11}, compression failure \cite{Zurlo11}, band localisation \cite{Gei14},  wrinkles \cite{Liu16, Pelr00, DuPl06, Yang17, Su18, Bortot18}, membrane wrinkling \cite{Greaney19}, etc.

Various methods have been proposed for avoiding electric breakdown without sacrificing the large actuation.
For example, if the material is pre-stretched before the voltage is applied, electric breakdown may be avoided, but the stretch gain achieved might be reduced \cite{Su18c}.
Another method proposed is \emph{charge-controlled} actuation, as shown experimentally by Keplinger et  al.  \cite{Kepl10} and theoretically by Li et al. \cite{BoLi11}.
In charge-controlled actuation, charges of opposite signs are sprayed on opposite planar surfaces of a dielectric plate, inducing a potential difference, and hence an electric field in the dielectric, thereby inducing a deformation.
In principle this method of actuation annihilates the possibility of snap-through because the theoretical charge-stretch loading curves are monotonic \cite{BoLi11}.
In this paper, we investigate the stability of a charge-driven dielectric plate, which has not been considered previously and the results of which are significantly different from those for voltage control.

We first focus on equi-biaxial loading and show that charge-controlled actuation is stable since the Hessian criterion--or rather, its version for this problem--for onset of instability is never met (Section \ref{Hessian criterion in the equi-biaxial case}).
This result is far from straightforward to obtain, because the Hessian determinant of the energy density is always \emph{negative}, from which it could erroneously be concluded that the actuation is unstable. In fact, we show that the \emph{second variation of the free energy of the whole system is always positive}, which ensures stability throughout.
This is in sharp contrast to the corresponding situation for voltage-controlled actuation, which is well-known \cite{ZhSu07} to become unstable once a critical voltage is reached.

We then highlight another new, and complementary, feature of charge control by showing that \emph{charge-controlled actuation is also stable with respect to geometric instability} because, provided the material is pre-stretched, small-amplitude inhomogeneous wrinkled solutions superposed on the large homogeneous actuation do not develop (Section \ref{Geometric instability in the equi-biaxial case}).  Again, this contrasts with the situation for voltage-controlled actuation, for which dielectric plates eventually wrinkle under sufficiently large voltages \cite{Ogden2014,DO2019,Su18c}.

In Section \ref{Uni-axial pre-stretch by a weight} we model the experiments of Keplinger et al. \cite{Kepl10} where a plate was pre-stretched by a weight prior to charge-controlled actuation.
We thus study the stability of a homogeneously deforming plate under uni-axial tension and charge-actuation and again we find Hessian-based and geometric stability in this case,  again contrary to the corresponding situation for voltage-controlled actuation. 

Finally in Section \ref{Finite Element simulations} we use \emph{Finite Element simulations} to account for the finite dimensions of plates.
We find that in the equi-biaxial case there are no differences between the results of the homogeneous loading analytical modelling and
those of the Finite Element  method, because the plate is free to stretch laterally  and the loading curves are indeed monotonic (no snap-through).
However, for the uni-axial case we find that the clamping of the plate required to apply the weight leads to non-homogeneous deformations with local variations of stresses and strains compared to the homogeneous solution, and that these effects build up and eventually lead to a breakdown of the simulation.
We identify the point of breakdown as corresponding to the appearance of compressive stresses in the plate.


\section{Equations of electroelasticity\label{sec2}}


Consider the stress-free reference configuration $\mathcal{B}_r$ of an electroelastic material in the absence of an electric field and applied mechanical loads.
Points in $\mathcal{B}_r$ are labelled by the position vector $\vec{X}$.
When subject to loads and an electric field under static conditions the material occupies the configuration $\mathcal{B}$, with the material point $\vec{X}$ now at $\vec{x}$.
Let $\vec{F} =\Grad\vec{x}$ denote the deformation gradient from $\mathcal{B}_r$  to $\mathcal{B}$, where $\Grad$ is the gradient operator with respect to $\vec{X}$.
We denote by $\vec{E}$ and $\vec{D}$, respectively, the electric field and electric displacement vectors in $\mathcal{B}$, and by $\vec{\tau}$ the Cauchy stress tensor (which in general depends on $\vec{F}$ and either $\vec{E}$ or $\vec{D}$).

It has been found advantageous \cite{Ogden05, ZhSu07}  to formulate constitutive equations in terms of the Lagrangian field variables, denoted $\vec{E}_L$, $\vec{D}_L$, and the nominal stress tensor $\vec{T}$, which are related to $\vec{E}$, $\vec{D}$ and $\vec{\tau}$ by the following pull-back operations (from $\mathcal{B}$ to $\mathcal{B}_r$)
\begin{equation}
\vec{E}_L=\vec{F}^T\vec{E},
\qquad \vec{D}_L=J\vec{F}^{-1}\vec{D},
\qquad \vec{T}=J\vec{F}^{-1}\vec{\tau},\label{L-E}
\end{equation}
where $J=\det\vec{F}$.

The constitutive equations are based on the use of so-called `total' energy functions, depending either on $\vec{F}$ and $\vec{E}_L$, denoted $\Omega$, or on  $\vec{F}$ and $\vec{D}_L$, denoted $\Omega^*$, with the (partial) Legendre transform connection
\begin{equation}
\Omega^*(\vec{F},\vec{D}_L)=\Omega(\vec{F},\vec{E}_L)+\vec{D}_L\cdot\vec{E}_L.\label{OmegaOmegastar}
\end{equation}
Henceforth, we confine attention to incompressible materials, so that the constraint $J\equiv 1$ is in force.  Then we have the constitutive relations
\begin{equation}
\vec{T}=\frac{\partial \Omega}{\partial \vec{F}}-p\vec{F}^{-1},
\qquad
\vec{T}=\frac{\partial \Omega^*}{\partial \vec{F}}-p^*\vec{F}^{-1},\label{const-rels1}
\end{equation}
(where $p$ and $p^*$ are Lagrange multipliers associated with the constraint, in general with $p^*\neq p$), and
\begin{equation}
\vec{D}_L=-\frac{\partial\Omega}{\partial\vec{E}_L},
\qquad
\vec{E}_L=\frac{\partial\Omega^*}{\partial\vec{D}_L}.
\end{equation}

The governing equations are
\begin{equation}
\Div\; \vec{T}=\mathbf{0},
\qquad
\Curl\vec{E}_L=\mathbf{0},
\qquad
\Div\vec{D}_L=0,\label{governing}
\end{equation}
where $\Div$ and $\Curl$ are the divergence and curl operators with respect to $\vec{X}$.
We shall consider the situation in which there is no external field, so that on the boundary $\partial\mathcal{B}_r$ of $\mathcal{B}_r$ the standard electric boundary conditions associated with the equations \eqref{governing} are simply
\begin{equation}
\vec{T}^T\vec{N}=\vec{t}_A,
\qquad
\vec{N}\times \vec{E}_L=\mathbf{0},
\qquad
\vec{N}\cdot\vec{D}_L=-\sigma_F \quad\mbox{on}\  \partial\mathcal{B}_r,\label{BCS}
\end{equation}
where $\vec{N}$ is the unit outward normal on $\partial\mathcal{B}_r$, $\vec{t}_A$ is the applied mechanical traction per unit area of $\partial\mathcal{B}_r$ and $\sigma_F$ is the surface charge density per unit area of $\partial\mathcal{B}_r$.

In considering applications to dielectric elastomers, which are isotropic electroelastic materials, the functional dependence of $\Omega$ and $\Omega^*$ can be expressed in terms of five invariants.
First of all, the isotropic purely kinematic invariants defined by
\begin{equation}
I_1 = \tr \vec{c},
\qquad
I_2 = \tfrac{1}{2}[I_1^2-\tr (\vec{c}^2)], \label{invariants1}
\end{equation}
where $\vec{c}=\vec{F}^T\vec{F}$ is the right Cauchy--Green deformation tensor.  Secondly, invariants associated with $\vec{E}_L$, which typically are taken to be
\begin{equation}
I_4  = \vec{E}_L \cdot \vec{E}_L,
\qquad
I_5  = \vec{E}_L \cdot (\vec{c}^{-1} \vec{E}_L),
\qquad
I_6  = \vec{E}_L \cdot (\vec{c}^{-2} \vec{E}_L), \label{invariants2}
\end{equation}
as in \cite{Ogden05}, and, thirdly, invariants associated with $\vec{D}_L$, here defined by
\begin{equation}
I_4^*  = \vec{D}_L \cdot \vec{D}_L,
\qquad
I_5^*  = \vec{D}_L \cdot (\vec{c} \vec{D}_L),
\qquad I_6^*  = \vec{D}_L \cdot (\vec{c}^2\vec{D}_L), \label{invariants3}
\end{equation}
as used in \cite{Ogden05} in different notation.

The expanded forms of the constitutive relations \eqref{const-rels1}, when converted to Eulerian form using \eqref{L-E} with $J=1$, are
\begin{align}
& \vec{\tau}=2\Omega_1\vec{b}+2\Omega_2(I_1\vec{b}-\vec{b}^2)-p\vec{I}-2\Omega_5\vec{E}\otimes\vec{E}-2\Omega_6(\vec{b}^{-1}\vec{E}\otimes\vec{E}+\vec{E}\otimes\vec{b}^{-1}\vec{E}),\label{tau1}
\\
& \vec{\tau}=2\Omega_1^*\vec{b}+2\Omega_2^*(I_1\vec{b}-\vec{b}^2)-p^*\vec{I}+2\Omega_5^*\vec{D}\otimes\vec{D}+2\Omega_6^*(\vec{b}\vec{D}\otimes\vec{D}+\vec{D}\otimes\vec{b}\vec{D}),\label{tau2}
\\
& \vec{D}=-2(\Omega_4\vec{b}+\Omega_5\vec{I}+\Omega_6\vec{b}^{-1})\vec{E},
\label{D-E}
\\
& \vec{E}=2(\Omega_4^*\vec{b}^{-1}+\Omega_5^*\vec{I}+\Omega_6^*\vec{b})\vec{D},\label{DEDE}
\end{align}
where $\vec{I}$ is the identity tensor, $\vec{b}=\vec{F}\vec{F}^T$ is the left Cauchy--Green deformation tensor, $\Omega_i=\partial \Omega/\partial I_i,\, i=1,2,4,5,6$, $\Omega_i^*=\partial \Omega^*/\partial I_i,\,i=1,2$, and  $\Omega_i^*=\partial \Omega^*/\partial I_i^*,\, i=4,5,6$.


\subsection{Specialization to biaxial deformations of a plate}


We now consider the application of the above theory to the biaxial deformation of a rectangular plate.

The plate has sides of lengths $L_1,L_2,L_3$ in the reference configuration $\mathcal{B}_r$, where $L_2=H$ is the thickness of the plate, which is small compared with its lateral dimensions.
Mechanical loads are applied in the $1$ and $3$ directions; also, a potential difference, say $V$, exists between the  major surfaces of the plate and the associated charges on the surfaces are denoted $\pm Q$.
As a result, the plate is stretched homogeneously with stretches $\lambda_1$ and $\lambda_3$ parallel to the major surfaces, and, by incompressibility, a stretch $\lambda_2=\lambda_1^{-1}\lambda_3^{-1}$ normal to the major surfaces.
The potential difference generates an electric field with a single component $E=E_2$, associated with an electric displacement component $D=D_2$.
The corresponding components of the Lagrangian fields are $E_L=\lambda_2E$ and $D_L=\lambda_2^{-1}D$.

In terms of the potential difference $V$ and the associated charges $\pm Q$ on the surfaces, we have the simple connections
\begin{equation}
E_L=-V/H,\qquad D_L=-\sigma_F=-Q/L_1L_3.
\end{equation}
Thus, for a fixed potential, $E_L$ is fixed, while fixed charge $Q$ corresponds to fixed $D_L$.

For this combination of deformation and electric field, $\Omega$ and $\Omega^*$ specialize accordingly.
The invariants are now given in terms of the independent stretches $\lambda_1,\lambda_3$ and $E_L$ and $D_L$ by
\begin{align}
& I_1=\lambda_1^2+\lambda_3^2+\lambda_1^{-2}\lambda_3^{-2},
&& I_2=\lambda_1^{-2}+\lambda_3^{-2}+\lambda_1^2\lambda_3^2,
&& \label{invariants1a}
\\
&I_4=E_L^2,
&& I_5=\lambda_1^2\lambda_3^2E_L^2,
&& I_6=\lambda_1^4\lambda_3^4E_L^2, \label{invariants2a}
\\
& I_4^*=D_L^2,
&& I_5^*=\lambda_1^{-2}\lambda_3^{-2}D_L^2,
&& I_6^*=\lambda_1^{-4}\lambda_3^{-4}D_L^2. \label{invariants3a}
\end{align}
We denote the specializations of $\Omega$ and $\Omega^*$  by $\omega$ and $\omega^*$, respectively, and the independent variables by $(\lambda_1,\lambda_3,E_L)$ and $(\lambda_1,\lambda_3,D_L)$, respectively, with, from the connection \eqref{OmegaOmegastar},
\begin{equation}
\omega^*(\lambda_1,\lambda_3,D_L)=\omega(\lambda_1,\lambda_3,E_L)+D_LE_L.\label{omegaomegastar}
\end{equation}

Since the resulting deformation is purely biaxial the corresponding nominal stress is coaxial with the edges of the plate;
we denote its components by $t_1$, $t_2$, $t_3$.

We now assume that there is no mechanical traction on the major faces of the plate so that the boundary condition \eqref{BCS}$_1$ yields $t_2=0$.
Then, on elimination of the hydrostatic stress from \eqref{tau1} and \eqref{tau2}, we obtain the simple formulas
\begin{equation}
t_1=\frac{\partial\omega}{\partial \lambda_1}=\frac{\partial\omega^*}{\partial \lambda_1},\qquad t_3=\frac{\partial\omega}{\partial \lambda_3}=\frac{\partial\omega^*}{\partial \lambda_3},\label{t1t3}
\end{equation}
and from \eqref{DEDE}
\begin{equation}
D_L=-\frac{\partial\omega}{\partial E_L},\qquad E_L=\frac{\partial\omega^*}{\partial D_L}.\label{DLEL}
\end{equation}

The particular case of \emph{equi-biaxial deformations} is of special interest, for then, with $\lambda_1 = \lambda_3 = \lambda$, and incompressibility giving  $\lambda_2 = \lambda^{-2}$, we may introduce the following further specialisations of the total energy functions,
\begin{equation}
\tilde{\omega}(\lambda,E_L)=\omega(\lambda,\lambda,E_L),\qquad \tilde{\omega}^*(\lambda,D_L)=\omega^*(\lambda,\lambda,D_L).
\end{equation}
We also have  $t_1=t_3=t$, say,  so that
\begin{equation}
t=\frac{1}{2}\frac{\partial \tilde{\omega}}{\partial\lambda}=\frac{1}{2}\frac{\partial \tilde{\omega}^*}{\partial\lambda},\qquad D_L=-\frac{\partial \tilde{\omega}}{\partial E_L},\qquad E_L=\frac{\partial \tilde{\omega}^*}{\partial D_L}.
\end{equation}

For illustration, we now consider models for which $\vec{D}=\varepsilon\vec{E}$, where $\varepsilon$, the material permittivity, is taken to be a constant.
These are ``ideal'' dielectrics in the terminology of Suo \cite{Suo10}.
Note that this linear relationship has recently been verified \cite{ZuDe18} using experimental data  for low to moderate values of the electric field for the acrylic dielectric elastomer VHB 4905.
In general, $\varepsilon$ may depend on the deformation, as has been shown in \cite{WM2007}, for example, for the acrylic dielectric elastomer VHB 4910,
but for our present purposes we consider it to be a material constant.

Then, $\Omega$ and $\Omega^*$ have the forms
\begin{align}
& \Omega=W(I_1,I_2)-\frac{\varepsilon}{2} I_5=W(I_1,I_2)-\frac{\varepsilon}{2}\vec{E}_L\cdot(\vec{c}^{-1}\vec{E}_L),\label{modelE}
\\[12pt]
& \Omega^*=W(I_1,I_2)+\frac{1}{2\varepsilon}I_5^*=W(I_1,I_2)+\frac{1}{2\varepsilon}\vec{D}_L\cdot(\vec{c}\vec{D}_L),\label{modelD}
\end{align}
and, for the biaxial deformations of a plate considered above,
\begin{equation}
\omega=w(\lambda_1,\lambda_3)-\frac{\varepsilon}{2}\lambda_1^2\lambda_3^2E_L^2,\qquad \omega^*=w(\lambda_1,\lambda_3)+\frac{1}{2\varepsilon}\lambda_1^{-2}\lambda_3^{-2}D_L^2,
\end{equation}
where $w(\lambda_1,\lambda_3)=W(I_1,I_2)$ with $I_1$ and $I_2$ given by \eqref{invariants1a} and $D_L=\lambda_1^2\lambda_3^2E_L$.
For equi-biaxial deformations, we have
\begin{equation}
\tilde{\omega}=\tilde{w}(\lambda)-\frac{\varepsilon}{2}\lambda^4E_L^2,\qquad \tilde{\omega}^*=\tilde{w}(\lambda)+\frac{1}{2\varepsilon}\lambda^{-4}D_L^2,
\end{equation}
where $\tilde{w}(\lambda)=w(\lambda,\lambda)$ and $D_L=\varepsilon\lambda^4E_L$.

For our subsequent applications we consider two representative energy density functions, a \emph{neo-Hookean} dielectric model and a \emph{Gent}  dielectric model, defined, in the two representations, by
\begin{align} \label{gent}
& \Omega_{nH}  = \frac{\mu}{2}(I_1-3) - \frac{\varepsilon}{2}I_5,
&& \Omega_G = -\frac{\mu J_m}{2} \ln \left( 1 - \frac{I_1 -3}{J_m} \right) - \frac{\varepsilon}{2}I_5,\\[6pt]
& \Omega_{nH}^* = \frac{\mu}{2}(I_1-3) +\frac{1}{2\varepsilon}I_5^*,
&& \Omega_G^* = -\frac{\mu J_m}{2} \ln \left( 1 - \frac{I_1 -3}{J_m} \right) + \frac{1}{2\varepsilon}I_5^*,
\end{align}
where $\mu$ is the shear modulus in the absence of an electric field and $J_m$ is a stiffening parameter.
Notice that $\Omega_G$ recovers $\Omega_{nH}$ and $\Omega_G^*$ recovers $\Omega_{nH}^*$  in the limit $J_m \to \infty$.

We now express the equations in dimensionless form by defining the following quantities
\begin{align}
& \bar{\omega} = \omega/\mu,
&& \bar{\omega}^*=\omega^*/\mu,
&& \hat{\omega}=\tilde{\omega}/\mu,
&& \hat{\omega}^*=\tilde{\omega}^*/\mu,
\notag\\
&D_0= D_L/\sqrt{\mu\varepsilon},
&& E_0=E_L\sqrt{\varepsilon/\mu},
&& s=t/\mu,
&&
\end{align}
so that $D_0=\lambda^4E_0$ in the equi-biaxial case, and
\begin{equation}
s = \frac{1}{2}\frac{\partial \hat{\omega}}{\partial \lambda}= \frac{1}{2}\frac{\partial \hat{\omega}^*}{\partial \lambda},\qquad D_0 = - \frac{\partial \hat{\omega}}{\partial E_0},\qquad
E_0 =  \frac{\partial \hat{\omega}^*}{\partial D_0}. \label{s_general}
\end{equation}

Based on either $\hat{\omega}$ or $\hat{\omega}^*$, we now obtain the expression for $D_0$ in terms of $\lambda$ and $s$ for the neo-Hookean and  Gent dielectric models as
\begin{equation}
D_0 = \sqrt{\lambda^6 -1 -\lambda^5 s},\qquad D_0= \sqrt{\frac{\lambda^6-1}{1-(2\lambda^2 + \lambda^{-4}-3)/J_m} - \lambda^5 s}, \label{loading_gent}
\end{equation}
respectively, and note that the latter reduces to the former when $J_m \rightarrow\infty$.

\begin{figure}[!h]
\centering
\includegraphics[width=\textwidth]{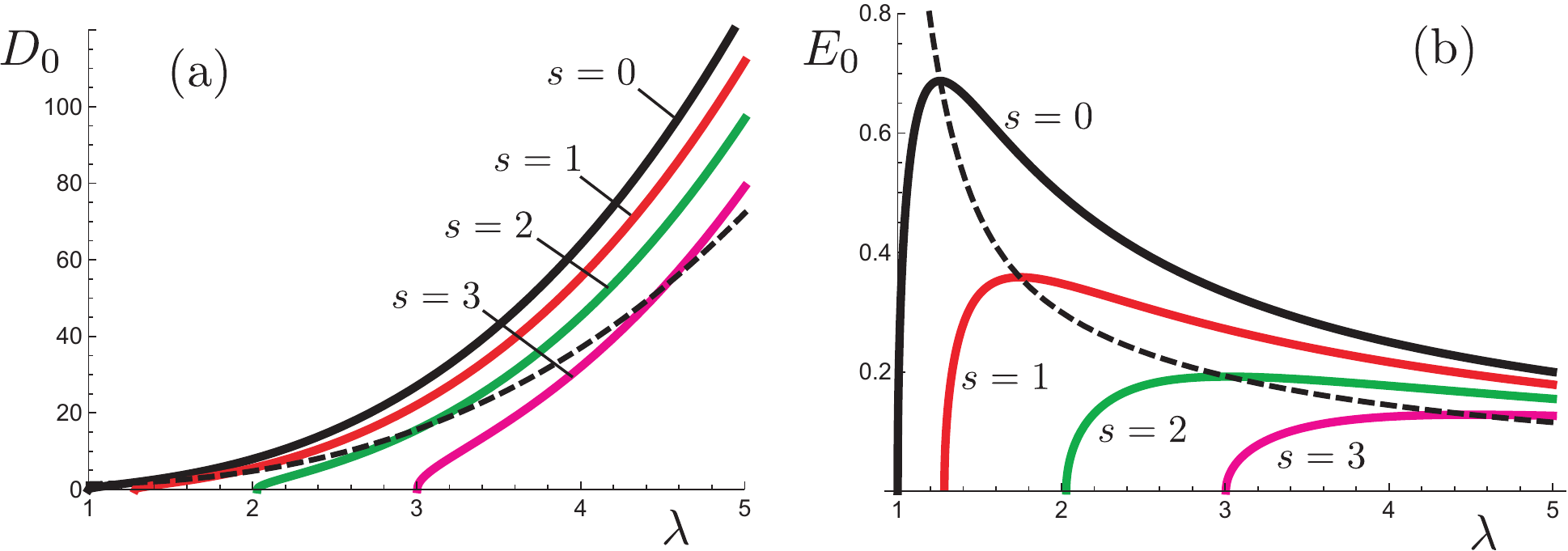}
\caption{Plots of (a) $D_0$ versus $\lambda$ and (b) $E_0$ versus $\lambda$ based on equation \eqref{loading_gent}$_1$ and the connection $E_0=\lambda^{-4}D_0$ for the neo-Hookean dielectric in equi-biaxial deformation, for values of non-dimensional pre-stress $s=0,1,2,3$ (continuous curves).
In (a) we also display the (dashed) curve of $D_0=\sqrt{(\lambda^6+5)/3}$, the intersections of which with the continuous curves correspond to the maxima in (b).\label{nh-d-e}}
\end{figure}

\begin{figure}[!h]
\centering
\includegraphics[width=\textwidth]{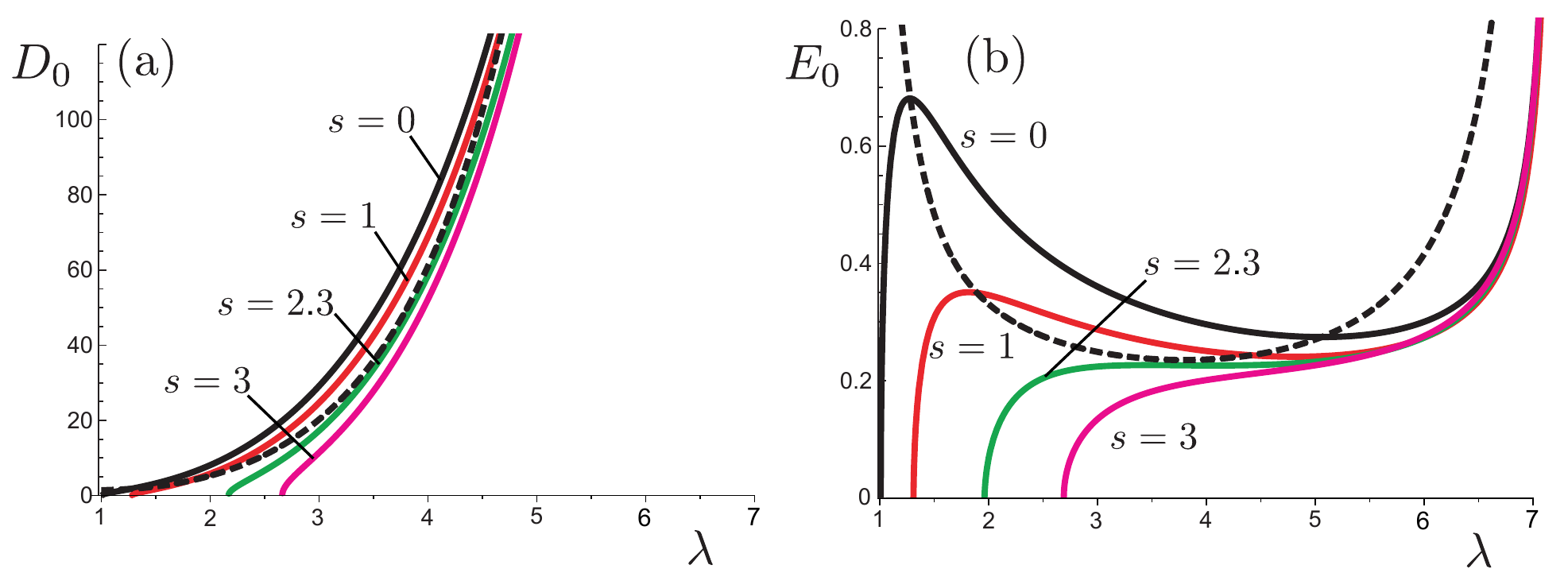}
\caption{Plots of (a) $D_0$ versus $\lambda$ and (b) $E_0$ versus $\lambda$ based on equation \eqref{loading_gent}$_2$ and $E_0=\lambda^{-4}D_0$ for the Gent model, for fixed values of $s=0,1,2.3,3$ (continuous curves) and in (a) the (dashed) curve of $D_0$ versus $\lambda$, the intersections of which with the continuous curves correspond to the maxima in (b). Note that for  larger values of $s$ there is no intersection.  Note also that the value $s=2.3$ has been used here instead of $s=2$ to enable the dashed curve to be distinguished from the continuous curve at larger values of $\lambda$ in (a).\label{gent-d-e}}
\end{figure}

Figures \ref{nh-d-e}(a) and \ref{gent-d-e}(a) show plots of these curves with $D_0$ versus $\lambda$ for several fixed values of $s$, and  Figures \ref{nh-d-e}(b) and \ref{gent-d-e}(b) display the corresponding plots of $E_0$ versus $\lambda$ based on the connection $E_0=\lambda^{-4}D_0$.
The value $J_m=97.2$ given by Gent  \cite{Gent96} has been used here.
Also shown in Figure \ref{nh-d-e}(a)  is the curve $D_0=\sqrt{(\lambda^6+5)/3}$, which cuts the fixed $s$ curves at points where $E_0$ is a maximum in Figure \ref{nh-d-e}(b), which also shows the corresponding dashed curve.
Similarly for the Gent dielectric in Figures \ref{gent-d-e}(a) and \ref{gent-d-e}(b), although, for the larger values of $s$, there is no maximum in (b) and no corresponding intersection.

We now turn to the analysis of the stability of the plate based on the Hessian criterion.

\subsection{Analysis of the Hessian stability criterion}
\label{Hessian criterion in the equi-biaxial case}


\emph{Electro-mechanical instability} is often considered to occur when the Hessian matrix associated with the second variation of the free energy for the whole system ceases to be positive definite \cite{ZhSu07}.
The rationale of this criterion is that equilibrium corresponds to an extremum of the free energy (and thus its first variation is zero), and that the equilibrium is stable when it corresponds to a minimum of the free energy (and then its second variation is positive).

In different notation and in dimensionless form, the free energy of the whole system, here denoted $\psi^*$, considered in \cite{ZhSu07} has the form
\begin{equation}
\psi^*(\lambda_1,\lambda_3,D_0)=\bar{\omega}^*(\lambda_1,\lambda_3,D_0)-s_1\lambda_1-s_3\lambda_3-D_0E_0,
\end{equation}
and vanishing of its first variation (for fixed $s_1,s_3,E_0$), with $s_1=t_1/\mu$, $s_3=t_3/\mu$, yields the dimensionless versions of the constitutive relations involving $\omega^*$ in \eqref{t1t3} and \eqref{DLEL}.

If, instead, we use $E_0$ as the independent electric variable, then the corresponding `energy', denoted $\psi$, vanishing of the first variation of which yields the constitutive relations in terms of $\omega$  in \eqref{t1t3} and \eqref{DLEL}, is given by
\begin{equation}
\psi(\lambda_1,\lambda_3,E_0)=\bar{\omega}(\lambda_1,\lambda_3,E_0)-s_1\lambda_1-s_3\lambda_3+D_0E_0.
\end{equation}

Note that, on use of \eqref{omegaomegastar} in dimensionless form, we have $\psi^*=\psi-D_0E_0$, so that $\psi$ is the Legendre transform of $\psi^*$ with respect to the conjugate variables $E_0$ and $D_0$ related by \eqref{s_general}$_4$.

For the free energy $\psi^*$ of the whole system to be at a minimum, its second variation must be positive, i.e. the associated Hessian matrix must be positive definite, at a point of equilibrium.
The second variations of $\psi^*$ and $\psi$ are written compactly as
\begin{equation}
\delta^2\psi^*=\delta\mathbf{a^*}\vec{\cdot}\left(\vec{\mathcal{H}^*}\delta\mathbf{a^*}\right),\qquad
\delta^2\psi=\delta\mathbf{a}\vec{\cdot}\left(\vec{\mathcal{H}} \, \delta\mathbf{a}\right),
\label{2nd-variation}
\end{equation}
respectively, with first variations  $\delta\mathbf{a}^*=[\delta\lambda_1,\delta\lambda_3,\delta D_0]^T$, $\delta\mathbf{a}=[\delta\lambda_1,\delta\lambda_3,\delta E_0]^T$, where $\vec{\mathcal{H}^*}$ and $\vec{\mathcal{H}}$ are the corresponding \emph{Hessian matrices}, which are given by
\begin{equation}
\vec{\mathcal{H}^*}=\left(\begin{array}{ccc}
\bar{\omega}_{11}^*&\bar{\omega}_{13}^*&\bar{\omega}_{1D_0}^*\\
\bar{\omega}_{13}^*&\bar{\omega}_{33}^*&\bar{\omega}_{3D_0}^*\\
\bar{\omega}_{1D_0}^*&\bar{\omega}_{3D_0}^*&\bar{\omega}_{D_0D_0}^*\end{array}\right),
\qquad
\vec{\mathcal{H}}=\left(\begin{array}{ccc}
\bar{\omega}_{11}&\bar{\omega}_{13}&\bar{\omega}_{1E_0}\\
\bar{\omega}_{13}&\bar{\omega}_{33}&\bar{\omega}_{3E_0}\\
\bar{\omega}_{1E_0}&\bar{\omega}_{3E_0}&\bar{\omega}_{E_0E_0}\end{array}\right),
\end{equation}
with the subscripts representing partial derivatives.

For the \emph{equi-biaxial case} these become $2\times 2$ matrices, given by
\begin{equation}
\vec{\mathcal{H}^*}=\left(\begin{array}{cc}
\hat{\omega}_{\lambda\lambda}^*&\hat{\omega}_{\lambda D_0}^*\\
\hat{\omega}_{\lambda D_0}^*&\hat{\omega}_{D_0D_0}^*\end{array}\right),
\qquad
\vec{\mathcal{H}}=\left(\begin{array}{cc}
\hat{\omega}_{\lambda\lambda}&\hat{\omega}_{\lambda E_0}\\
\hat{\omega}_{\lambda E_0}&\hat{\omega}_{E_0E_0}\end{array}\right),
\end{equation}
and we now focus on this case for illustration.

It is straightforward to show that $\hat{\omega}^*_{D_0D_0}=-1/\hat{\omega}_{E_0E_0}$ by using the formulas \eqref{s_general}$_{3,4}$.
Now the determinants of the Hessians above are given by
\begin{equation}
\det\vec{\mathcal{H}^*}=\hat{\omega}_{\lambda\lambda}^*\hat{\omega}_{D_0D_0}^*-\hat{\omega}_{\lambda D_0}^{*2},
\qquad
\det\vec{\mathcal{H}}=\hat{\omega}_{\lambda\lambda}\hat{\omega}_{E_0E_0}-\hat{\omega}_{\lambda E_0}^2,
\end{equation}
and on specializing \eqref{omegaomegastar} we have $\hat{\omega}^*(\lambda,D_0)=\hat{\omega}(\lambda,E_0)+D_0E_0$, from which the following connections, given in \cite{DO2019} in dimensional form, can be obtained:
\begin{equation}
  \det\vec{\mathcal{H}^*}=\hat{\omega}_{\lambda\lambda}\hat{\omega}_{D_0D_0}^*=-\hat{\omega}_{\lambda\lambda}/\hat{\omega}_{E_0E_0},
  \qquad
\det\vec{\mathcal{H}}=\hat{\omega}_{\lambda\lambda}^*\hat{\omega}_{E_0E_0}.
\end{equation}
These equations are independent of the specific forms of $\hat{\omega}^*$ and $\hat{\omega}$, and so are valid for any choice of (equi-biaxial) energy density function. 
 They have some interesting interpretations, which we now discuss in respect of the neo-Hookean dielectric, for which
 \begin{equation}
 \hat{\omega}^*=\tfrac{1}{2}(2\lambda^2+\lambda^{-4}-3)+\tfrac{1}{2}\lambda^{-4}D_0^2,
 \qquad
  \hat{\omega}=\tfrac{1}{2}(2\lambda^2+\lambda^{-4}-3)-\tfrac{1}{2}\lambda^4E_0^2,
 \end{equation}
 and hence
 \begin{equation}
 s=\tfrac{1}{2}\hat{\omega}_{\lambda}=\lambda-\lambda^{-5}-\lambda^3E_0^2=\lambda-\lambda^{-5}-\lambda^{-5}D_0^2=\tfrac{1}{2}\hat{\omega}_{\lambda}^*,
 \end{equation}
 and
 \begin{align}
 &  \hat{\omega}_{\lambda\lambda}^*=2(1+5\lambda^{-6}+5\lambda^{-6}D_0^2),
 && \hat{\omega}_{\lambda\lambda}=2(1+5\lambda^{-6}-3\lambda^2E_0^2),
\\
&  \hat{\omega}_{\lambda D_0}^*=-4\lambda^{-5}D_0,\quad \hat{\omega}_{D_0D_0}^*=\lambda^{-4},
&& \hat{\omega}_{\lambda E_0}=-4\lambda^3E_0,\quad \hat{\omega}_{E_0E_0}=-\lambda^4.
 \end{align}
 Thus here,
 \begin{equation}
 \det\vec{\mathcal{H}^*}=2\lambda^{-10}(\lambda^6+5-3D_0^2),
 \qquad
 \det\vec{\mathcal{H}}=-2\lambda^4(1+5\lambda^{-6}+5\lambda^2E_0^2).
 \end{equation}

 Note that the maxima of $E_0$ in Figure  \ref{nh-d-e}(b) correspond to $ \det\vec{\mathcal{H}^*}=0$, equivalently $\hat{\omega}_{\lambda\lambda}=0$, which also corresponds to a maximum of $s$ at fixed $E_0$.
 Note that $\vec{\mathcal{H}^*}$ is positive definite up to the maxima as $E_0$ is increased from $0$.
 By contrast, $s$ is monotonic with respect to $\lambda$ at fixed $D_0$ and $\hat{\omega}^*_{\lambda\lambda}>0$, while $\det\vec{\mathcal{H}}<0$ and $\vec{\mathcal{H}}$ is indefinite, thus defining a saddle point of $\hat{\omega}$.
 Note that it would be incorrect to conclude here that the charge-controlled actuation is unstable, as we now show.

Consider the connection
 \begin{equation}
\tilde{\omega}^*(\lambda,D_0)=\tilde{\omega}(\lambda,E_0)+E_0D_0,
\label{connection}
 \end{equation}
the first variation of which yields
 \begin{equation}
\tilde{\omega}^*_{\lambda}\delta\lambda+\tilde{\omega}^*_{D_0}\delta D_0=\tilde{\omega}_{\lambda}\delta \lambda+\tilde{\omega}_{E_0}\delta E_0+E_0\delta D_0+D_0 \delta E_0,
 \end{equation}
leading to
 \begin{equation}
\tilde{\omega}^*_{\lambda}=\tilde{\omega}_{\lambda},\qquad E_0=\tilde{\omega}^*_{D_0},\qquad D_0=-\tilde{\omega}_{E_0}.\label{equi-const-connects}
 \end{equation}
If we now take the second variation then the terms involving $\delta^2\lambda$, $\delta^2E_0$, $\delta^2D_0$ cancel and we are left with the quadratic connection
 \begin{equation}
\tilde{\omega}^*_{\lambda\lambda}(\delta\lambda)^2+2\tilde{\omega}^*_{\lambda D_0}\delta\lambda \delta D_0+\tilde{\omega}^*_{D_0D_0}(\delta D_0)^2 =
\tilde{\omega}_{\lambda\lambda}(\delta\lambda)^2+2\tilde{\omega}_{\lambda E_0}\delta\lambda \delta E_0+\tilde{\omega}_{E_0E_0}(\delta E_0 )^2 + 2\delta E_0 \delta D_0.\label{sec-var-equi}
 \end{equation}
From \eqref{equi-const-connects}$_3$ we obtain
 \begin{equation}
\delta D_0=-(\tilde{\omega}_{\lambda E_0}\delta\lambda+\tilde{\omega}_{E_0E_0}\delta E_0),
 \end{equation}
 and hence, by substituting for $\delta D_0$ on the right-hand side of \eqref{sec-var-equi}, we obtain
 \begin{equation}
\tilde{\omega}^*_{\lambda\lambda}(\delta\lambda)^2+2\tilde{\omega}^*_{\lambda D_0}\delta\lambda \delta D_0+\tilde{\omega}^*_{D_0D_0}(\delta D_0 )^2 = \tilde{\omega}_{\lambda\lambda}(\delta\lambda)^2 -
\tilde{\omega}_{E_0E_0}(\delta E_0)^2.\label{sec-var-equi2}
 \end{equation}

For stability we require the left-hand side to be positive since this is the second variation of the actual free energy $\psi^*$ (so the free energy is minimized), whether we have voltage-control of the deformation (when $\lambda$ and $D_0$ are free to vary) or charge-control of the deformation (when $\lambda$ and $E_0$ are free to vary).

For fixed $E_0$, in a \emph{voltage-controlled experiment}, this reduces simply to $\tilde{\omega}_{\lambda\lambda}>0$, and this fails where $E_0$ is a maximum.
For the neo-Hookean dielectric, it reads $\lambda^{-2}+5\lambda^{-8}-3E_0^2>0$, and $E_0=\sqrt{(\lambda^{-2} + 5\lambda^{-8})/3}$ is the plot going through the maxima of each loading curve for different values of the pre-load $s$, as shown by the dashed curve in Figure \ref{nh-d-e}(b).

For fixed $D_0$, in a \emph{charge-controlled experiment}, the left-hand side is positive if $\tilde{\omega}_{\lambda\lambda}^*>0$,
and for the neo-Hookean dielectric, this reads $1+5\lambda^{-6}(1+D_0^2)>0$, which holds true for all $D_0$.  In the case of a perfect dielectric, we have $D_0=\lambda^4E_0$, and hence $0=4\lambda^3\delta\lambda E_0+\lambda^4 \delta E_0$, so for the right-hand side of \eqref{sec-var-equi2} to be positive we have
 \begin{equation}
\tilde{\omega}_{\lambda\lambda}-16\tilde{\omega}_{E_0E_0}E_0^2\lambda^{-2}=2(1+5\lambda^{-6}+5\lambda^2E_0^2)>0,
 \end{equation}
which confirms the result  $\tilde{\omega}_{\lambda\lambda}^*>0$ for the neo-Hookean dielectric, and thus, that the second variation of the free energy is always positive. 

We can therefore conclude that \emph{under charge control, equi-biaxial activation is stable} according to the Hessian criterion since we have $\tilde{\omega}_{\lambda\lambda}^*>0$ for the considered neo-Hookean model.
On the other hand, \emph{activation under voltage control can become unstable} in the Hessian criterion sense, as is well known, since the inequality $\tilde{\omega}_{\lambda\lambda}>0$ can fail.  The results for the Gent model (not developed here) follow the same pattern.


\section{Incremental stability analysis}
\label{Geometric instability in the equi-biaxial case}


To investigate the possibility of \emph{geometric instabilities}, namely the formation of small-amplitude wrinkles on the faces of the plate, we linearise the governing equations and boundary conditions in the neighbourhood of a large deformation and initial electric field.

We introduce the incremental mechanical displacement $\vec{u}$, the incremental nominal stress tensor $\vec{\dot{T}}$ and the incremental Lagrangian electric field and displacement, $\vec{\dot{E}}_L$ and $\vec{\dot{D}}_L$, respectively, all of which are functions of the deformed position $\vec{x}$ \cite{Ogden2014}.  Let $\vec{\dot{T}}_0$, $\vec{\dot{E}}_{L0}$  and $\vec{\dot{D}}_{L0}$ denote their push-forward forms from the reference to the deformed configuration, as defined by $\vec{\dot{T}}_0=\vec{F\dot{T}}$, $\vec{\dot{E}}_{L0}=\vec{F}^{-T}\vec{\dot{E}}_{L}$, $\vec{\dot{D}}_{L0}=\vec{F\dot{D}}_{L}$.  These satisfy the governing equations
\begin{equation}
\div \;\vec{\dot{T}}_0=\mathbf{0},\qquad \curl \vec{\dot{E}}_{L0}=\mathbf{0},\qquad \div\vec{\dot{D}}_{L0}=0,
\end{equation}
and the relevant incremental constitutive equations are
\begin{equation}
\vec{\dot{T}}_0=\vec{\mathcal{A}}_0\vec{L}+p \vec{L}-\dot{p}\vec{I}+\vec{\mathbb{A}}_0\vec{\dot{E}}_{L0},\qquad \vec{\dot{D}}_{L0}=-\vec{\mathbb{A}}_0^T\vec{L}-\vec{\mathsf{A}}_0\vec{\dot{E}}_{L0},
\end{equation}
where $\vec{\mathcal{A}}_0, \vec{\mathbb{A}}_0$ and $\vec{\mathsf{A}}_0$ are, respectively, fourth-, third- and second-order electroelastic moduli tensors (see \cite{Su18} for their general expressions), and $\vec{L}$ is the displacement gradient $\grad \vec{u}$, $\vec{u}$ being the incremental displacement, which, by incompressibility, satisfies $\tr\vec{L}\equiv\div\vec{u}=0$.

Attention is now focused on two-dimensional wrinkles \cite{Su18} so that the fields are functions of the components $x_1,x_2$ of $\vec{x}$ only, and  $u_3 = \dot{E}_{L03} = \dot{D}_{L03} = 0$.  The governing equations then reduce to
\begin{equation}
\dot{T}_{011,1}+\dot{T}_{021,2}=0,\quad \dot{T}_{012,1}+\dot{T}_{022,2}=0, \quad \dot{E}_{L01,2}-\dot{E}_{L02,1}=0,\quad \dot{D}_{L01,1}+\dot{D}_{L02,2}=0,\label{2Deqs}
\end{equation}
where subscripts $1$ and $2$ following a comma signify differentiation with respect to $x_1$ and $x_2$, respectively.

From \eqref{2Deqs}$_3$ we can introduce the scalar electric potential $\varphi$ such that
\begin{equation}
	\dot{E}_{L01}  = -\varphi_{,1}, \qquad \dot{E}_{L02}  = -\varphi_{,2}.
\end{equation}

We now focus on models of the form
\begin{equation}
\Omega(I_1,I_5)=W(I_1)-\tfrac{1}{2}\varepsilon I_5,
\end{equation}
for which the relevant components of the moduli tensors reduce to
\begin{align}
& \mathcal{A}_{01111}= 4W_{11}\lambda_1^4+2W_1\lambda_1^2,
&& \mathcal{A}_{02222}=4W_{11}\lambda_2^4+2W_1\lambda_2^2-3\varepsilon E_2^2,
\notag \\[0.5ex]
& \mathcal{A}_{01122}=4W_{11}\lambda_1^2\lambda_2^2-\varepsilon E_2^2,
&& \mathcal{A}_{01221}=\mathcal{A}_{02112}=0,
\notag \\[0.5ex]
& \mathcal{A}_{01212}=2W_{1}\lambda_1^2-\varepsilon E_2^2,
&& \mathcal{A}_{02121}=2W_1\lambda_2^2,
\notag \\[0.5ex]
& \mathbb{A}_{012|1}=\mathbb{A}_{021|1}=\varepsilon E_2,
&&\mathbb{A}_{022|2}=2\varepsilon E_2,
\notag \\[0.5ex]
& \mathbb{A}_{011|1}=\mathbb{A}_{022|1}=\mathbb{A}_{011|2}=0,
&& \mathbb{A}_{012|2}=\mathbb{A}_{021|2}=0,
\notag \\[0.5ex]
& \mathsf{A}_{011}=\mathsf{A}_{022}=-\varepsilon,
&& \mathsf{A}_{012}=0.
\end{align}
Note that we used the connection $p=\mathcal{A}_{02121}$, which is a special case of a general formula given in, for example, equation (9.88) of \cite{Ogden2014}.
The vertical bar between the components of $\mathbb{A}_0$ is used to distinguish the single index (associated with a vector) from the pair of indices associated with a second-order tensor.

Now, for brevity, we introduce the notations
\begin{equation}
a=\mathcal{A}_{01212},
\qquad
2b=\mathcal{A}_{01111}+ \mathcal{A}_{02222}-2 \mathcal{A}_{01122},
\qquad
c= \mathcal{A}_{02121},
\qquad
d=\mathbb{A}_{012|1}.
\end{equation}
Then, on elimination of $\dot{p}$ and use of the incompressibility equation $u_{1,1}+u_{2,2}=0$, the required incremental constitutive equations can be written compactly in the form
\begin{align}
& \dot{T}_{011}=\dot{T}_{022}+2(b+c) u_{1,1}+2d\varphi_{,2},
&&
\notag\\[0.5ex]
& \dot{T}_{012}=au_{2,1}+cu_{1,2}-d\varphi_{,1},
&&
\dot{T}_{021}=c(u_{1,2}+u_{2,1})-d\varphi_{,1},
\notag \\[0.5ex]
&\dot{D}_{L01}=-d(u_{1,2}+u_{2,1})-\varepsilon \varphi_{,1},
&& \dot{D}_{L02}=2du_{1,1}-\varepsilon \varphi_{,1}. \label{con-eqns}
\end{align}

We now convert the system of equations to a first-order system with six variables based on the Stroh approach.
For this purpose we choose the variables $u_1$, $u_2$, $\varphi$, $\dot{T}_{021}$, $\dot{T}_{022}$, $\dot{D}_{L02}$ and consider increments that are sinusoidal in the $x_1$ direction, i.e. solutions of the form
 \begin{equation}
	\left\lbrace u_1, u_2, \varphi, \dot{T}_{021}, \dot{T}_{022}, \dot{D}_{L02} \right\rbrace = \Re \lbrace e^{\mathrm{i}kx_1} \left[ U_1, U_2, \Phi, \mathrm{i}k\Sigma_{21},\mathrm{i}k\Sigma_{22}, \mathrm{i}k\Delta \right] \rbrace,
\end{equation}
where $U_1$, $U_2$, $\Phi$, $\Sigma_{21}$, $\Sigma_{22}$ and $\Delta$ are all functions of $kx_2$, $k = 2\pi / \mathcal L$ is the wave number and $\mathcal{L}$ is the wavelength of the wrinkles.

We now arrange the variables so that they all have the same dimensions by defining a Stroh vector $\vec{\eta}$ as
\begin{equation}
\vec{\eta}=(\vec{U},\vec{S})=[U_1, U_2, \sqrt{\varepsilon/\mu}\,\Phi, \Sigma_{21}/\mu,\Sigma_{22}/\mu,\Delta/\sqrt{\mu\varepsilon}],\label{strohvector}
\end{equation}
where $\vec{U}$ is the `displacement' vector and $\vec{S}$ is the `traction' vector.
After a little manipulation, the equations \eqref{2Deqs} and \eqref{con-eqns} are cast in the form
\begin{align}
&\eta_1'=\mathrm{i}(-\eta_2+\bar{d}\bar{c}^{\,-1}\eta_3+\bar{c}^{\,-1}\eta_4),
\notag \\
&\eta_2'=-\mathrm{i}\eta_1,
\notag \\
&\eta_3'=\mathrm{i}(2\bar{d}\eta_1-\eta_6),
\notag \\
& \eta_4'=\mathrm{i}[-(2\bar{b}+2\bar{c}+4\bar{d}^2)\eta_1-\eta_5+2\bar{d}\eta_6],
\notag \\
&\eta_5'=\mathrm{i}[(\bar{c}-\bar{a})\eta_2-\eta_4],
\notag \\
& \eta_6'=\mathrm{i}[(\bar{d}^2\bar{c}^{\,-1}+1)\eta_3+\bar{d}\bar{c}^{\,-1}\eta_4],
\end{align}
where $\bar{a}=a/\mu$, $\bar{b}=b/\mu$, $\bar{c}=c/\mu$ and $\bar{d}=d/\sqrt{\mu\varepsilon}$, and a prime denotes differentiation with respect to $kx_2$  .
Similarly to Su et al. \cite{Su18} we can thus write the equations in Stroh form, i.e. as
\begin{equation}
	\vec{\eta} ' = \mathrm{i} \vec{N\eta}, \label{stroh-form}
\end{equation}
where $\vec{N}$ is the Stroh matrix and $\vec{\eta}$ is the Stroh vector, defined in \eqref{strohvector}.
Note that the vector $\vec{\eta}$ is different from its counterpart in the voltage-controlled case \cite{Su18}, due to the different electric boundary conditions and scalings.
In the voltage-controlled case, the incremental electric boundary condition is in terms of the electric potential $\Phi$ (which must be zero on the faces), whereas in the charge-controlled case, the incremental electric boundary condition is in terms of the electric displacement $\Delta$.
In the present situation the Stroh matrix has the dimensionless form
\begin{equation}
\vec{N} = \left[ \begin{matrix} \vec{N}_1 & \vec{N}_2   \\
\vec{N}_3 & \vec{N}_1^T\end{matrix} \right],
\end{equation}
where
\begin{align}
	\vec{N}_1 &= \left[ \begin{matrix} 0            & -1 & \bar{d}/\bar{c} \\
			-1           & 0  & 0           \\
			2\bar{d} & 0  & 0
			\end{matrix} \right], & \vec{N}_2 &= \left[ \begin{matrix} 1/\bar{c} & 0 & 0           \\
			0           & 0 & 0           \\
			0           & 0 &-1\end{matrix} \right], & \vec{N}_3 &= \left[ \begin{matrix} -2(b+c)-4\bar{d}^2 & 0   & 0                \\
			0                    & \bar{c}-\bar{a} & 0                \\
			0                    & 0   & \bar{d}^2/\bar{c} +1\end{matrix} \right].
\end{align}

For the models \eqref{modelE} and \eqref{modelD} for which $W(I_1,I_2)$ depends on only $I_1$, i.e. $W = W(I_1)$, including the Gent dielectric model \eqref{gent}, in equi-biaxial activation we have
\begin{equation}
	\bar{a}  =  2\lambda^2 \bar{W}' - \lambda^{-4} D_0^2, \qquad
	\bar{c}  = 2 \lambda^{-4} \bar{W}', \qquad
	2\bar{b} = 4 (\lambda^{-4}-\lambda^2)^2\bar{W}''+\bar{a}+\bar{c}, \qquad
	\bar{d} = \lambda^{-2} D_0,\label{gent-moduli}
\end{equation}
where $\bar{W}(I_1)=W(I_1)/\mu$, and henceforth we restrict attention to this specialization.  For the Gent dielectric,
\begin{equation}
	\bar{W}'   = \frac{1}{2\left[1 - (2\lambda^2+\lambda^{-4}-3)/J_m\right]},\quad
	               \bar{W}''  = \frac{1}{2J_m\left[1 - (2\lambda^2+\lambda^{-4}-3)/J_m\right]^2}, \label{W}
\end{equation}
and we recall that $E_0 = \lambda^{-4}D_0$.
For the neo-Hookean dielectric, the expressions simplify considerably as: $\bar{W}' = 1/2$ and $\bar{W}''=0$.

To investigate the conditions for wrinkling to occur, it is sufficient to calculate the thin-plate and thick-plate limits of the dispersion equation, as the behaviour of a plate with finite thickness lies in between the two \cite{Su18}.

The \emph{thin-plate limit} is calculated from the Stroh matrix as \cite{Shuv00,Su18}
\begin{equation}
	\det \vec{N}_3 = 0,
\end{equation} which simplifies here to
\begin{equation}
	(\bar{a}-\bar{c})(\bar{b}+\bar{c}+2\bar{d}^{\,2})(\bar{d}^{\,2}+\bar{c})=0.
	\label{detN3}
\end{equation}
As in the voltage-controlled case, the thin-plate limit can be separated into symmetric and anti-symmetric modes.
Anti-symmetric modes are governed by the equation $\bar{a}-\bar{c}=0$, as in the voltage-controlled case.
For the neo-Hookean and the Gent dielectric models this yields
\begin{equation}
D_0 = \sqrt{\lambda^6-1}, \qquad
	D_0 = \sqrt{\frac{\lambda^6-1}{1-(2\lambda^2 + \lambda^{-4}-3)/J_m}}, \label{antisym_thin}
\end{equation}
respectively, which is the same as \eqref{loading_gent} in the absence of pre-stress ($s=0$).
No symmetric modes are possible as they are governed by the equation $\bar{b}+\bar{c}+2\bar{d}^2 = 0$, which has no real solutions in $(\lambda, D_0)$.
Likewise, the third factor in \eqref{detN3} yields no solutions.

To calculate the \emph{thick-plate limit}, we first construct a matrix with the eigenvectors $\vec{\eta}^{(j)}$, $j=1,2,3$, of $\vec N$ with corresponding eigenvalues with positive imaginary part, stacked as the columns as follows
\begin{equation}
	\left[ \begin{matrix}
			\vec{A} \\ \vec{B}
		\end{matrix} \right] = \left[ \begin{matrix} |                & |                & |                \\
			\vec{\eta}^{(1)} & \vec{\eta}^{(2)} & \vec{\eta}^{(3)} \\
			|                & |                & |\end{matrix} \right],
\end{equation}
where $\vec{A}$ and $\vec{B}$ defined above are $3\times3$ matrices and explicit expressions for the components of  $\vec{\eta}^{(j)}$, $j=1,2,3$, are given as follows
 \begin{align}
\vec{\eta}^{(1)} &= \left[ \begin{array}{c} 0 \\ 0 \\ 1 \\ -\lambda^{-2} D_0 \\ -\mathrm{i}\lambda^{-2} D_0 \\ -\mathrm{i} \end{array} \right], &
\vec{\eta}^{(j)} &= \left[ \begin{array}{c} \mathrm{i}\lambda^8 p_j \\ -\lambda^8 \\ \lambda^6 D_0 \\ -2\lambda^4 \bar{W}' (p_j^2 +1) -\lambda^4 D_0^2 \\ -2\mathrm{i} \bar{W}' p_j^{-1} \lambda^4 (\lambda^6 + p_j^2)  \\ \mathrm{i}\lambda^6 p_j D_0 \end{array} \right],  \label{eigenvectors}
\end{align}
for $j=2,3$ and where $p_{2,3}$ and $\bar{W}'$ are given by \eqref{evalue} below and \eqref{W}$_1$, respectively.

Then the thick-plate limit is given by
\begin{equation}
	\det\left(\mathrm{i}\vec{BA}^{-1}\right) = 0.\label{detStroh}
\end{equation}
Based on the analysis of Stroh (see Ting \cite{Ting96} or Shuvalov \cite{Shuv00}, for instance),  we recall that $\mathrm{i}\vec{BA}^{-1}$ is Hermitian and the above equation is a single real equation, as distinct from $\det\vec{B}=0$, which is a complex equation, although its real and imaginary parts are in proportion.

For  models with $W = W(I_1)$, including that of Gent, equation \eqref{detStroh} is a quadratic in $D_0^2$, explicitly 
\begin{equation} 
D_0^4 -2 \bar{W}' \left[ \lambda^3 (p_2 + p_3) +2(\lambda^3-1) \right] D_0^2 -4 \bar{W}^{\prime\,2} \left[ \lambda^3 ( p_2 + p_3 )^2 - (\lambda^3-1)^2 \right] =0,
\label{short-wave}
\end{equation}
where 
\begin{equation}
p_{2,3} = \frac{\lambda^3 +1}{2} \sqrt{1 + 2(\lambda - \lambda^{-2})^2 \frac{\bar{W}''}{\bar{W}'}} \mp \frac{\lambda^3 -1}{2} \sqrt{1 + 2(\lambda + \lambda^{-2})^2 \frac{\bar{W}''}{\bar{W}'}},
	\label{evalue}
\end{equation} where $(-)$ and $(+)$ correspond to $p_2$ and $p_3$, respectively.
Note that for the neo-Hookean specialization, since $\bar{W}' = 1/2$, $\bar{W}'' = 0$, we obtain $p_2 = 1$ and $p_3 = \lambda^3$ and the thick-plate limit becomes
\begin{equation}
	D_0^4 - (\lambda^6+3\lambda^3-2)D_0^2 -(\lambda^9 + \lambda^6 +3\lambda^3 -1) =0. \label{ideal-shortwave}
\end{equation}
In the absence of charge ($D_0=0$), this reduces to the classical elastic case and recovers the critical stretch for surface instability under equi-biaxial stretch of Green and Zerna \cite{GrZe54}, specifically $\lambda = 0.666$.

\begin{figure}[!h]
	\centering
	\includegraphics[width=0.9\textwidth]{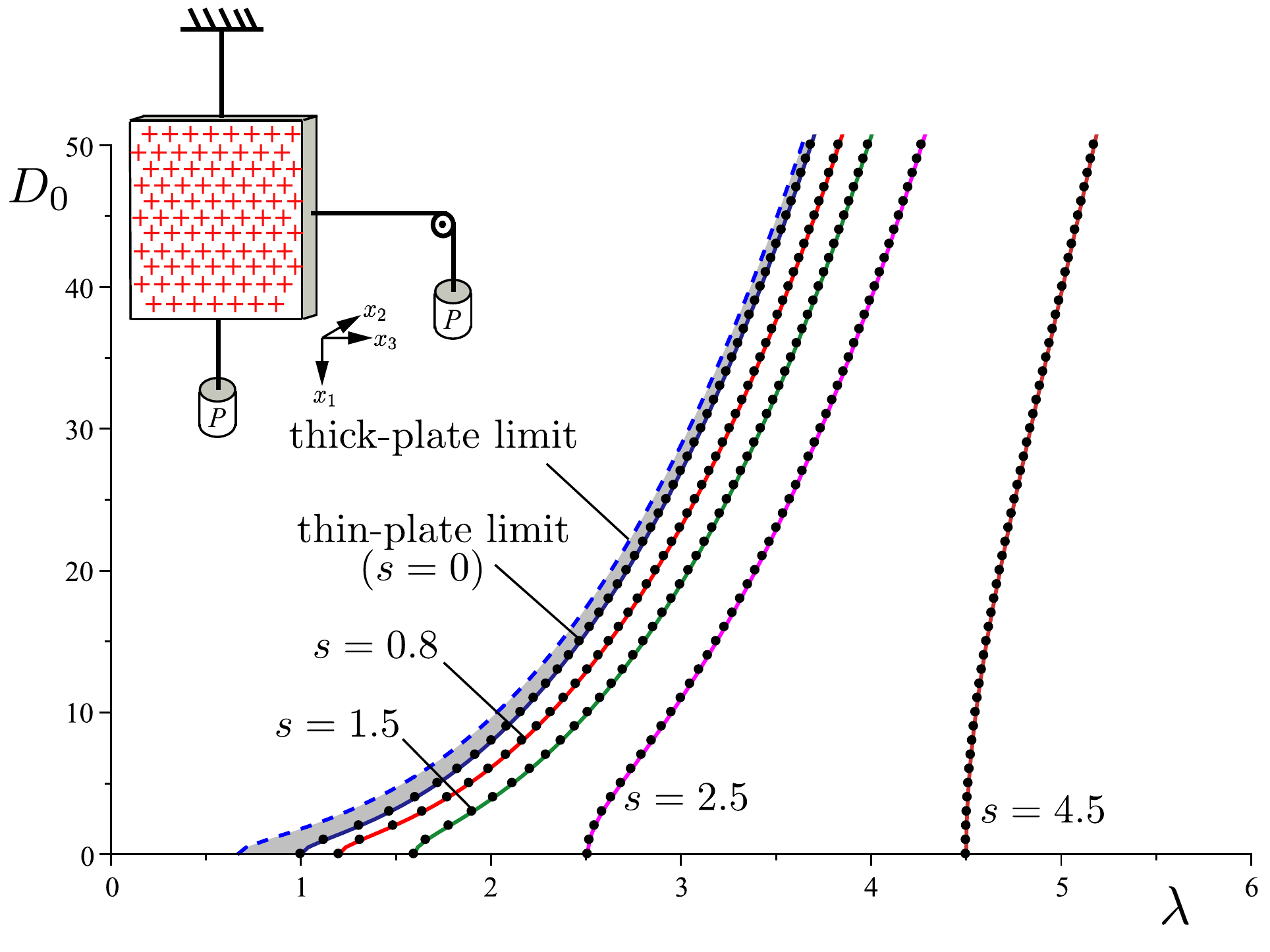}
	\caption{Wrinkles are not expressed in equi-biaxially pre-stretched charge-controlled plates.
	Here the solid curves are the loading curves for the neo-Hookean dielectric model with pre-stresses $s=0, 0.8, 1.5, 2.5, 4.5$.
	The dashed curve is the thick-plate limit \eqref{ideal-shortwave}. None of the pre-stretched curves cross the greyed zone where wrinkling occurs, between the thick-plate (dashed curve) and thin-plate ($s=0$ loading curve) limits, so wrinkling does not take place.
	The dots are the result of Finite Element calculations using COMSOL Multiphysics{\small \textregistered} (Section \ref{Finite Element simulations}), which turn out to be  very stable numerically.
		We conducted the same calculations for the Gent dielectric  with $J_m = 97.2$ and found almost identical plots (not shown here).}
	\label{gent-equi-biaxial}
\end{figure}

We plot the thick- and thin-plate limits, along with the loading curves \eqref{loading_gent}$_1$ for the neo-Hookean dielectric model
for different values of pre-stress in Figure \ref{gent-equi-biaxial}. The loading curves are monotonic, and so the material will \emph{not} experience the snap-through phenomenon of voltage-controlled actuation \cite{BoLi11}. 
As shown in the previous section, this is connected to to the sign of the second variation of the free energy being always positive. 

These theoretical predictions are compared with Finite Element simulations (see Section \ref{Finite Element simulations}), the results of which are represented by dots in the figure, which also exhibit the stability.

The region between the thick-plate and thin-plate limits is where wrinkling could occur. However, the pre-stretched loading curves do not cross this region, so there is no wrinkling. Charge-controlled dielectric plates are therefore \emph{geometrically stable}, and will not exhibit wrinkling (provided $s>0$).
This situation again contrasts with voltage-controlled plates, which can wrinkle in compression, as here, but also in extension  \cite{Ogden2014,DO2019,Su18c}, which is not possible here.


\section{Activation under uni-axial dead load}
\label{Uni-axial pre-stretch by a weight}


In order to model the experiments of Keplinger et al. \cite{Kepl10}, we now consider a plate that is pre-stretched by a uni-axial dead load.
A weight is applied in the $x_1$-direction and  charges on the lateral faces of the dielectric so that an electric field is induced in the $x_2$-direction.

In dimensionless form, the loading curves relating the uni-axial stress $s$, the electric displacement component $D_0$ and the electric field $E_0$ to the stretches $\lambda_1$ and $\lambda_3$  for the neo-Hookean model \eqref{gent}$_1$ are given by
\begin{align}
	s= \lambda_1 - \lambda_1 ^{-1} \lambda_3^2, \qquad
	D_0^2 = \lambda_1 ^2 \lambda_3 ^4 -1, \qquad E_0^2=\lambda_1^{-2}-\lambda_1^{-4}\lambda_3^{-4},\label{uni-axial-loading}
\end{align}
which lead to expressions for  $D_0$--$\lambda_1$ and $E_0$--$\lambda_1$ relationships in terms of $s$ (see, for example, \cite{Lu12, HuaSu12} for details in the voltage-controlled case), namely
\begin{equation}
	D_0 = \sqrt{\lambda_1^4(\lambda_1-s )^2-1},\qquad E_0=\lambda_1^{-1}\sqrt{1-\lambda_1^{-4}(\lambda_1-s)^{-2}}.\label{DEuniax}
\end{equation}

\begin{figure}[!h]
\centering
\includegraphics[width=\textwidth]{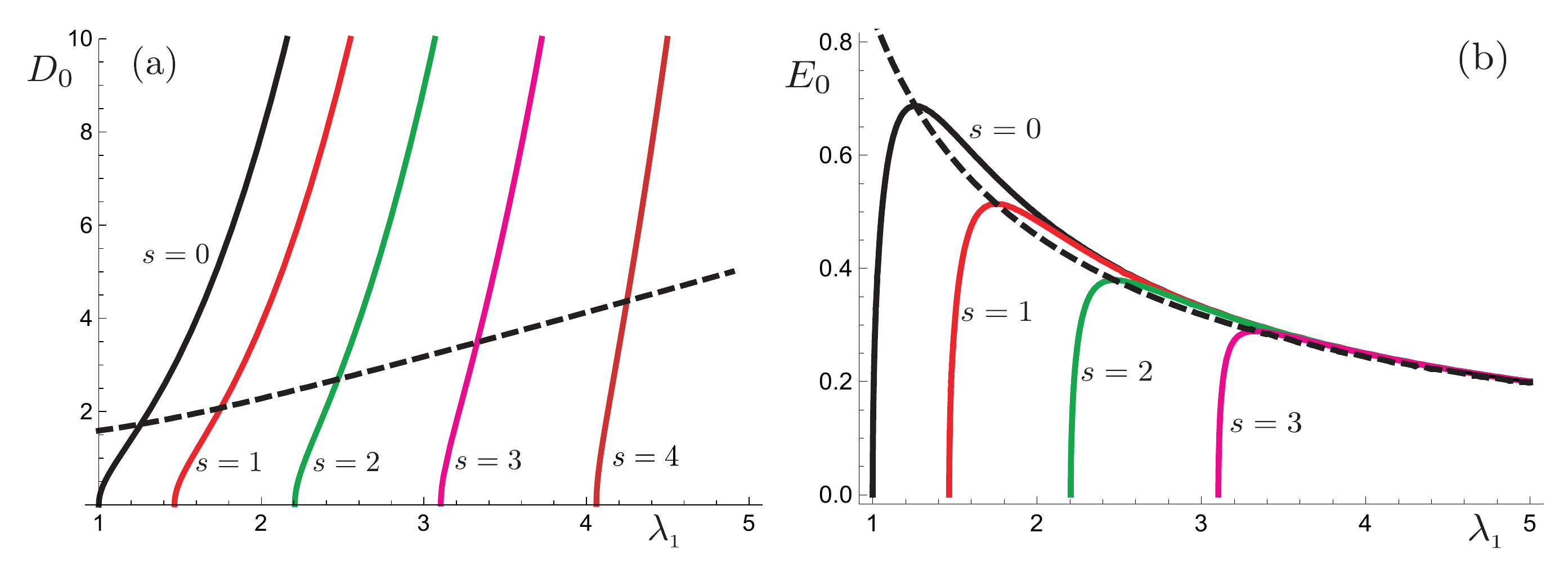}
\caption{Plots of (a) $D_0$  versus $\lambda_1$ for $s=0,1,2,3,4$, and (b) $E_0$ versus $\lambda_1$ for $s=0,1,2,3$ based on the equations in  \eqref{DEuniax} for the neo-Hookean model.  In each case a plot (dashed) of $\det\boldsymbol{\mathcal{H}}^*=0$ in terms of (a) $D_0$  versus $\lambda_1$, and (b) $E_0$  versus $\lambda_1$ is included.\label{figde}}
\end{figure}

Plots of $D_0$ and $E_0$ versus $\lambda_1$ based on \eqref{DEuniax} for several fixed values of $s$ are shown in Figures \ref{figde}(a) and  \ref{figde}(b), respectively, as the continuous curves.  Notice, in particular, that $D_0$ is monotonic in $\lambda_1$, while $E_0$ exhibits maxima, these behaviours being associated with loss of Hessian stability, as we elaborate on below.

It is a simple matter to extend the problem of \emph{minimizing the free energy} $\psi^*$ associated with the whole system from the equi-biaxial to the uni-axial case in order to study material stability.
First, we note that for the general biaxial case the second variations of the connection \eqref{connection}, corresponding to \eqref{sec-var-equi} and \eqref{sec-var-equi2} in the equi-biaxial case,  are
\begin{align}
 \bar{\omega}^*_{11}\delta\lambda_1^2  + 2\bar{\omega}^*_{13}\delta\lambda_1\delta\lambda_3 & + \bar{\omega}^*_{33}\delta\lambda_3^2 +2\bar{\omega}^*_{1D_0}\delta\lambda_1 \delta D_0+2\bar{\omega}^*_{3D_0}\delta\lambda_3 \delta D_0+\bar{\omega}^*_{D_0D_0}\delta D_0^2\notag\\[4pt]
& =\bar{\omega}_{11}\delta\lambda_1^2+2\bar{\omega}_{13}\delta\lambda_1\delta\lambda_3+\bar{\omega}_{33}\delta\lambda_3^2
\notag\\
& \qquad \qquad +2\bar{\omega}_{1E_0}\delta\lambda_1 \delta E_0+2\bar{\omega}_{3E_0}\delta\lambda_3 \delta E_0 +\bar{\omega}_{E_0E_0}\delta E_0^2+2\delta E_0\delta D_0\notag\\[4pt]
&=\bar{\omega}_{11}\delta\lambda_1^2+2\bar{\omega}_{13}\delta\lambda_1\delta\lambda_3+\bar{\omega}_{33}\delta\lambda_3^2-\bar{\omega}_{E_0E_0}\delta E_0^2.
\label{equiv}
\end{align}

For the second variations of the free energy of the whole system $\psi^*$ to be positive, the $3 \times 3$ Hessian matrix $\boldsymbol{\mathcal{H}^*}$ must be positive definite (recall \eqref{2nd-variation}$_1$).
According to the equality above, this is equivalent under voltage control (when $\lambda_1$, $\lambda_3$ and $D_0$ are free to vary and $E_0$ is fixed) to
\begin{equation}
\bar{\omega}_{11}\delta\lambda_1^2+2\bar{\omega}_{13}\delta\lambda_1\delta\lambda_3+\bar{\omega}_{33}\delta\lambda_3^2 >0,
\end{equation}
for non-zero $\delta\lambda_1$ and/or $\delta\lambda_3$, i.e. it is equivalent to  the leading $2\times 2$ minor in  $\boldsymbol{\mathcal{H}}$ being positive definite.

On the other hand, under charge control (when $\lambda_1$, $\lambda_3$ and $E_0$ are free to vary and $D_0$ is fixed), the left hand side of the equality \eqref{equiv} tells us that  leading $2\times 2$ minor of $\boldsymbol{\mathcal{H}^*}$ should be positive definite for stability, i.e.
\begin{equation}
\bar{\omega}^*_{11}\delta\lambda_1^2+2\bar{\omega}^*_{13}\delta\lambda_1\delta\lambda_3+\bar{\omega}^*_{33}\delta\lambda_3^2 >0,
\end{equation}
for non-zero $\delta\lambda_1$ and/or $\delta\lambda_3$.

The latter inequality always holds for the neo-Hookean dielectric model, since $\bar{\omega}^*_{11}>0$  and
the leading $2\times 2$ minor of $\boldsymbol{\mathcal{H}^*}$ is positive definite, with determinant
\begin{equation}
1+3\lambda_1^{-4}\lambda_3^{-4}(\lambda_1^2+\lambda_3^2) (1+D_0^2)+5\lambda_1^{-6}\lambda_3^{-6}(1+D_0^2)^2,
\end{equation}
which factorizes in the equi-biaxial case, with $\lambda_1=\lambda_3=\lambda$, as
\begin{equation}
[1+5\lambda^{-6}(1+D_0^2)][1+\lambda^{-6}(1+D_0^2)],
\end{equation}
the first factor coinciding with the corresponding result in the purely equi-biaxial case.

Also, using \eqref{DEuniax}$_2$, we find
\begin{equation}
\det\boldsymbol{\mathcal{H}^*}=4\lambda_1^{-6}\lambda_3^{-6}(3\lambda_3^2+\lambda_1^2-\lambda_1^2\lambda_3^6),
\end{equation}
which corresponds to
\begin{equation}
\bar{\omega}_{11}\delta\lambda_1^2+2\bar{\omega}_{13}\delta\lambda_1\delta\lambda_3+\bar{\omega}_{33}\delta\lambda_3^2 =0,
\end{equation}
for fixed $E_0$.
This condition means that the leading $2\times 2$ minor of $\boldsymbol{\mathcal{H}}$ is indefinite,
which can hold for fixed $E_0$ (at least for the neo-Hookean model), and we also have $\det\boldsymbol{\mathcal{H}}<0$.

Plots of $E_0$ versus $\lambda_1$ for the uni-axial case are shown in Figure \ref{figde}(b) for $s=0,1,2,3$, and the connection between $E_0$ and $\lambda_1$ where $\det\boldsymbol{\mathcal{H}^*}=0$ is also shown as the dashed curve that passes through the maximum points of  $E_0$.  In Figure \ref{figde}(a) are shown corresponding plots of $D_0$ (for $s=0,1,2,3,4$) versus $\lambda_1$, the dashed curve corresponding to where $\det\boldsymbol{\mathcal{H}^*}=0$.

In conclusion, for a neo-Hookean dielectric subject to a uni-axial dead load, activation with voltage control can become unstable, but charge-controlled  activation is always stable in the sense of the Hessian free energy criterion.

For the study of the \emph{geometric stability}, we again refer to the limit cases.  First, the \emph{thin-plate limit}, again $\det \vec N_3 = 0$, reduces to
\begin{equation}
	D_0^2 = \lambda_1^4 \lambda_3^2 -1.\label{thin-uni-axial}
\end{equation}
The \emph{thick-plate limit}, is a quadratic in $D_0^2$ given by
 \begin{equation}
	D_0^4 - \left( \lambda_1^4 \lambda _3^2 +3\lambda_1^2 \lambda_3 -2 \right)D_0^2 - \left( \lambda_1^6 \lambda_3^3 + \lambda_1^4 \lambda_3^2 + 3 \lambda_1^2 \lambda_3 -1 \right) =0. \label{thick-uni-axial}
\end{equation}
Note that these two equations apply for all $\lambda_1\,(>0)$  and, in particular, they recover the conditions for the equi-biaxial case \eqref{antisym_thin}$_1$ and \eqref{ideal-shortwave} when $\lambda_1=\lambda_3=\lambda$.

\begin{figure}[!h]
	\centering
	\includegraphics[width=0.9\textwidth]{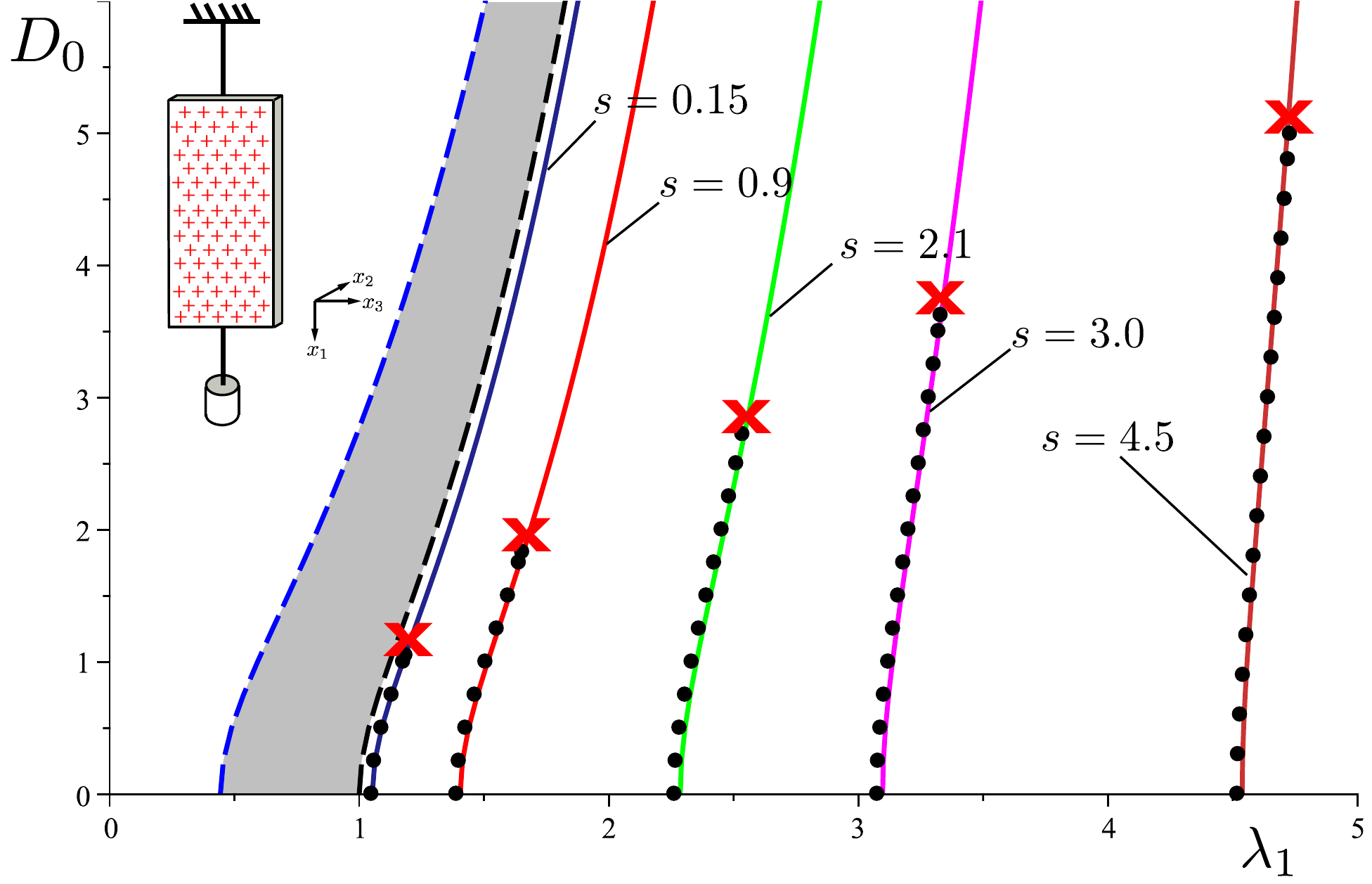}
	\caption{Wrinkles are not expressed for uni-axially-loaded, charge-driven dielectric plates.
	The solid curves are the loading curves for the neo-Hookean dielectric \eqref{gent}$_1$ with pre-stress $s= \alpha mg/(\mu A)$, where $\alpha = 0.05, 0.3, 0.7,1.0,1.5$, and the other characteristics taken from the Keplinger et al. \cite{Kepl10} membrane ($m= 150$ g, $\mu = 9833.07$ Pa, $A=50$ mm$^2$).
	The left-most dashed (blue) curve is the thick-plate limit curve \eqref{thick-uni-axial} and the other dashed (black) curve is the thin-plate limit \eqref{thin-uni-axial} curve, equivalent to the hypothetical no-weight curve ($s =0$).
	The shaded region between the thick and thin-plate limits represents values of $D_0$ and $\lambda_1$ for which wrinkling could occur. Because the loading curves for the pre-stressed plate ($s > 0$) are all monotonic, they will not cross into the wrinkling region, provided the material is pre-stretched, and so wrinkling will not occur in the direction of the uni-axial load.
	The dots result from Finite Element computations, and follow the theoretical curves closely, although the clamping of the plate creates local, non-homogeneous fields.
	The main difference with the theoretical predictions is that the simulations eventually breakdown numerically, as indicated by red crosses.}
	\label{uni-axial-ideal}
\end{figure}

The limit conditions above relate to wrinkles aligned with the direction of the uni-axial load.
In Figure \ref{uni-axial-ideal} we plot the corresponding $D_0$--$\lambda_1$ curves by solving each condition \eqref{thin-uni-axial} and \eqref{thick-uni-axial}
together with \eqref{uni-axial-loading}$_{2}$.   The loading curves \eqref{DEuniax}$_1$ are also plotted, for different values of uni-axial pre-stress $s$.

As in the equi-biaxial case, the thin-plate limit is equivalent to the loading curve in the absence of pre-stress ($s=0$).
The wrinkling zone between the thin- and thick-plate limits is not reached by any of the curves corresponding to a pre-stretch ($s>0$), and so the uni-axially pre-stretched plate will not wrinkle in the direction of the load.   Note that in the absence of charge ($D_0=0$), we again recover the purely elastic case, where $\lambda_2 = \lambda_1 ^{-1/2}$, and the critical stretch for uni-axial surface instability is the Biot value $\lambda_1 = 0.444$ \cite{Biot63}.

We  also investigated wrinkles perpendicular to the direction of the load using the same method.
There we looked for wrinkles in the $(x_2, x_3)$-plane, and constructed the Stroh formulation for the variables
\begin{equation}
	\left\lbrace u_3, u_2, \varphi, \dot{T}_{023}, \dot{T}_{022}, \dot{D}_{L02} \right\rbrace. \label{perpendicular-vars}
\end{equation}
We then found that the thin-plate condition is identical to the loading curve equation \eqref{uni-axial-loading}$_2$ for $s=0$, and that the thick-plate limit is
\begin{equation}
	D_0 ^4 - \left( \lambda_1^2 \lambda _3^4 +3\lambda_1 \lambda_3^2 -2 \right) D_0^2 - \left( \lambda_1^3 \lambda_3^6 + \lambda_1^2 \lambda_3^4 + 3 \lambda_1 \lambda_3^2 -1 \right) =0. \label{thick-plate23}
\end{equation}
On solving this condition together with \eqref{uni-axial-loading}$_2$, no real solutions are found, and so there are no wrinkles perpendicular to the uni-axial load.

In the next section we see that the Hessian and geometric stabilities found from the homogeneous deformation fields can be contradicted by local inhomogeneous effects, as shown in numerical simulations.


\section{Finite Element simulations}
\label{Finite Element simulations}


To complement the results of the theory, we developed  electroelastic Finite Element (FE) models of the equi-biaxial and the uni-axial experiments using the commercial software COMSOL Multiphysics{\small \textregistered} \cite{COMSOL}, and coupled the elasticity and electrostatics in two different ways.

In the fully coupled model, COMSOL{\small \textregistered} uses the second Piola--Kirchhoff stress tensor, denoted $\vec{P}$, and  implements  incompressibility via a volumetric energy function in the form $\kappa( \det \vec{F}-1)^2/2$, where $\kappa$ is the initial bulk modulus, taken to be orders of magnitude larger than the shear modulus.

The second way to solve the coupled problem is by considering the effect of the Maxwell stress tensor as a fictitious mechanical boundary condition in the purely elastic problem.
Since there are no charges within the volume of a dielectric, it is possible to consider the Maxwell stress as a pressure applied on the external faces of the volume.
This adds a boundary traction $\vec{\tau}_m \vec{n}$ to the mechanical problem, where $\vec{n}$ is the outward normal to the deformed surface of the specimen and $\vec{\tau}_m$ is the Maxwell stress tensor
\begin{equation}
	\vec{\tau}_m=\vec{E} \otimes \vec{D}-\tfrac{1}{2}(\vec{E}\vec\cdot\vec{D})\vec{I}.
\end{equation}

We found that both methods lead to the same results, although we noted that imposing the Maxwell stress tensor as a pressure boundary condition seemed to be a slightly more stable method in the uni-axial case.

In the equi-biaxial case, we found no difference between the predictions of the analytical model and those of the FE model, which also displayed stability and could be performed at any level of charge control, see Figure \ref{gent-equi-biaxial}.

\begin{figure}
	\centering
	\subfloat[][Stretch levels in the $x_1$-direction.]
	{\includegraphics[width=0.35\linewidth]{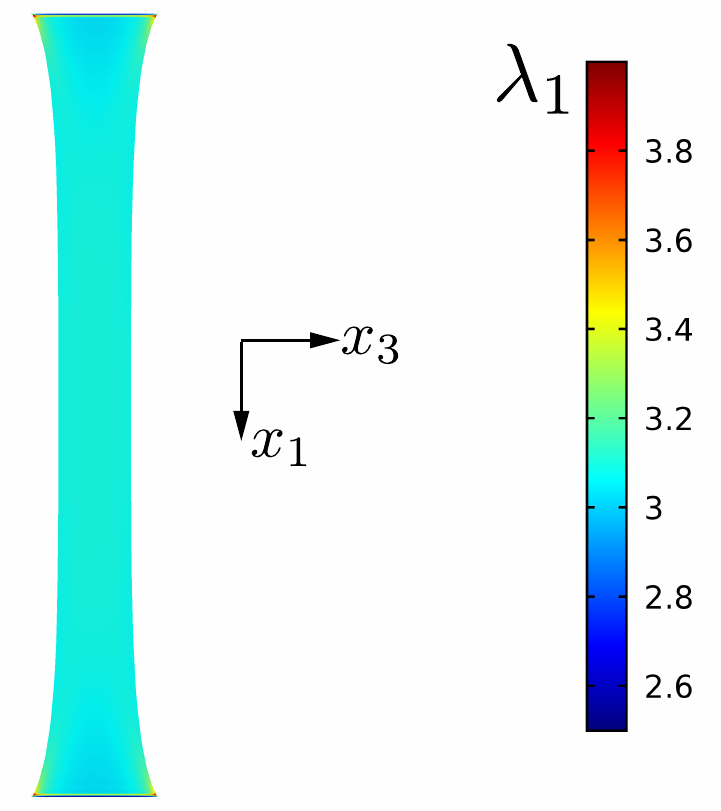}} \qquad\qquad\qquad
	\subfloat[][Stretch levels in the $x_3$-direction. ]
	{\includegraphics[width=0.35\linewidth]{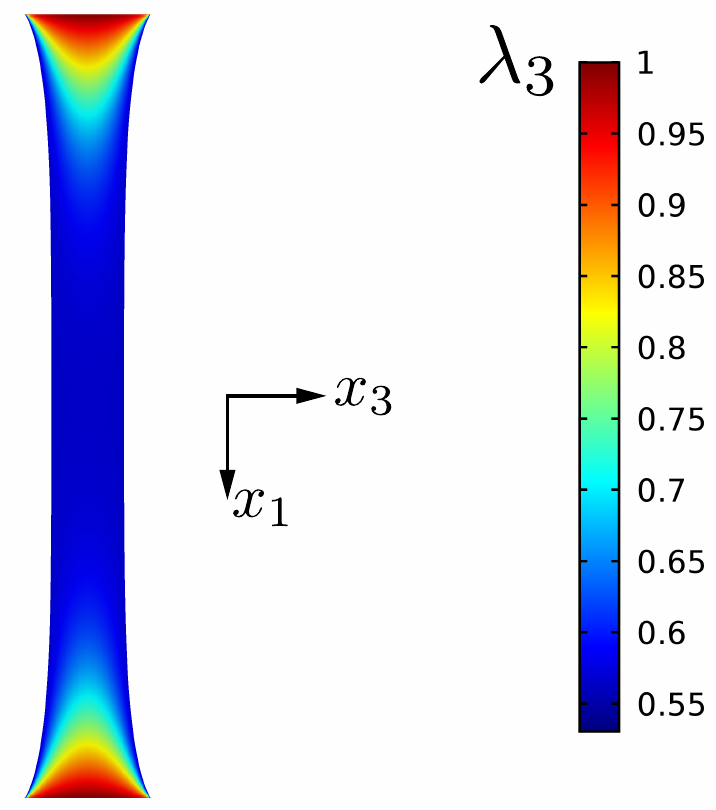}}
	\caption{Stretches in the dielectric plate after uni-axial loading by a weight, prior to activation, as computed by FE analysis using COMSOL Multiphysics{\small \textregistered}. We used the same physical characteristics as those in the experiments by Keplinger et al. \cite{Kepl10}.  Dimensions: length 100 mm; width 50 mm; thickness 1 mm. Attached  mass: 150 g.  Constitutive model: neo-Hookean dielectric with $\mu =$ 9833.07 Pa.}
	\label{fig:lambda1-3}
\end{figure}

By contrast, a major difference between the analytical model and the FE model arises in the uni-axial case, because the simulations for the latter eventually break down.
We identified the reason for this numerical breakdown to be due to the boundary conditions in the areas close to the clamping playing an initially  small but eventually significant role.
In the analytical model the strain is homogeneous and the material is free to deform in the transverse $x_3$-direction.
In the real-world experiments \cite{Kepl10} and in the FE numerical model, the top and bottom parts of the material are clamped and the strain is inhomogeneous in these neighbourhoods, see Figure~\ref{fig:lambda1-3}(b).
This behaviour is local, however, and the stretch in the direction of the uni-axial tension due to the weight (the $x_1$-direction) is almost completely homogeneous, as can be seen in Figure~\ref{fig:lambda1-3}(a).

\begin{figure}[!h]
	\centering
	\includegraphics[width=0.6\linewidth]{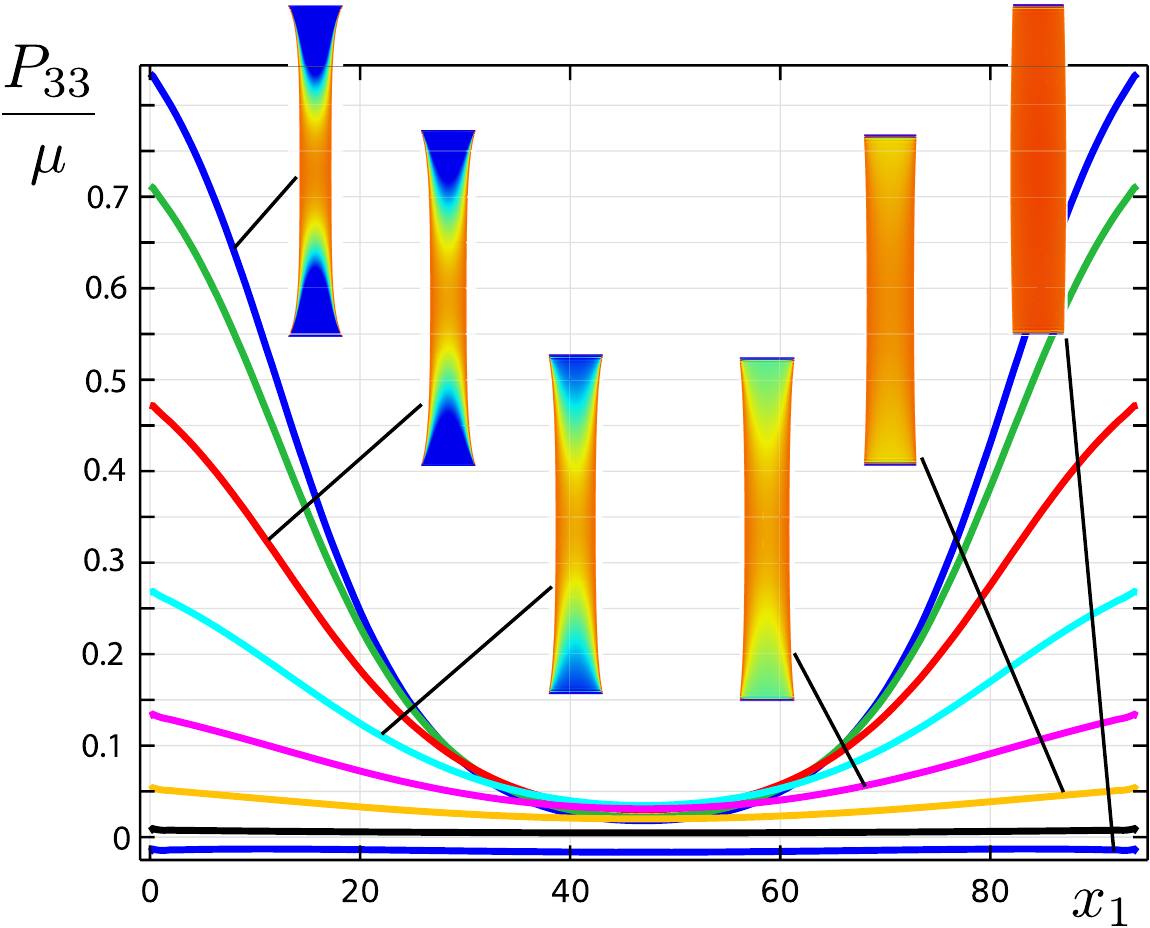}
	\caption{Finite Element simulations of a charge-driven plate subject to a dead-load. The total second Piola--Kirchhoff stress lateral component $P_{33}$ (normalised with respect to $\mu$) along the centre line of the material in the $x_1$-direction (direction of uni-axial tension, in cm).  The plate's characteristics are the same as for Figure \ref{fig:lambda1-3}.
The uppermost curve corresponds to the static dead-load condition ($D_0 = 0$);  then an increasing charge activation is performed until the simulation reaches the instability point, where the stress is slightly compressive throughout (lowest curve) and the computation crashes.  The colour coding for the levels of  $P_{33}$  in the simulations goes from about 1 kPa in blue to about 0 kPa in dark orange.
\label{fig:zs22}}
\end{figure}

When the stretched plate is electrically activated it expands in area and its thickness reduces. While it is free to expand in the direction of the dead load (the $x_1$-direction), the situation in the transverse $x_3$-direction is different.
There is a central zone where the influence of the clamping is weak, so that the normal stress component $P_{33}$ in the $x_3$-direction remains close to zero, as in the homogeneous case.
On the other hand, the portions of material closer to the clamping areas suffer from the fixed displacement in the $x_3$-direction imposed by the clamps.
There the application of the uni-axial tension due to the weight increases $P_{33}$, as is clearly visible in Figure~\ref{fig:zs22}.

When the plate is progressively activated with an increasing uniform charge distribution on its faces, the stress component $P_{33}$ near the clamping zone is progressively reduced until a critical value just below zero is reached: in this configuration, the plate undergoes a lateral compression that makes it buckle in the $x_3$-direction \cite{Zurlo11}.
At that point the FE computation breaks down, presumably because the stiffness matrix stops being positive definite and the solver has trouble converging.
This phenomenon does not occur in the analytical model and in the equi-biaxial case, as $\lambda_3$ is  homogeneous then and $P_{33}$ is imposed from the boundary condition and is identically equal to zero everywhere.

For larger weights, the levels of the $P_{33}$ stress component before activation are higher,  making it possible to activate the dielectric plate with a larger value of the electric charge before the instability condition is reached, as can be seen from the dots in  Figure~\ref{uni-axial-ideal}.

Despite the significant difference in the transverse behaviour between the analytical and numerical model, due to the different boundary condition imposed, there is very good agreement in the results, as can be seen in Figure~\ref{uni-axial-ideal}.   As long as the FE model stays below the point of negative $P_{33}$, the $D_0$--$\lambda$ curves follow those of the homogeneously deformed analytical model very closely.


\section{Conclusion}


In conclusion, we found that  both equi-biaxial and uni-axial modes in the charge-control actuation of a dielectric plate are stable, whether the stability analysis is based on a Hessian criterion for the free energy of the whole system, or on the formation of small-amplitude inhomogeneous wrinkles.

By comparing the different Hessian criteria that result from the voltage- and charge-control situations, we found that charge-controlled actuation is always stable with respect to the Hessian criterion, in complete contrast to voltage-controlled actuation, which, according to the Hessian criterion, can become unstable.

We also investigated the possibility of small-amplitude wrinkles and found that the wrinkling conditions in the limiting cases of thin and thick plates occur only in compression, whereas it has been shown that wrinkles can exist in extension in the voltage-controlled case \cite{Su18}. As a result, charge-controlled actuation, which always occurs in extension, is also geometrically stable, again in contrast to voltage-control actuation.

To account for the difference  between a theoretical homogeneous uni-axial deformation and the local inhomogeneous fields created by clamps in practice, we also conducted Finite Element simulations to verify our analytical results. We found complete agreement in the equi-biaxial case and very close agreement in the uni-axial case. So the assumption of homogeneous deformation is well justified for modelling the behaviour of charge-controlled activation of a dielectric plate in equi-biaxial stretch, and in uni-axial stretch when the aspect ratio of the specimen is high.

In the uni-axial case, Finite Element simulations reveal that the fringe effects are localised in a portion of area near the clamping zone and that they do not significantly affect the homogeneous loading curves of the system, although they have a strong effect on the eventual instability of the setup, a possibility that the homogeneous solution cannot capture.  We may also argue that the emergence of compressive lateral stresses inside the plate seen in the simulations has a real-world counterpart, and that an equi-biaxial pre-stress leads to larger actuations than a uni-axial pre-stress in practice.

Of course, the plate may become unstable due to other causes than free energy instability or inhomogeneous small-amplitude wrinkles.
Other mechanisms include for instance charge localisation  \cite{Lu14} or thickness effects \cite{Fu18, Zurlo17}.


\section*{Acknowledgements}

This work is supported by a Government of Ireland Postgraduate Scholarship from the Irish Research Council (Project GOIPG/2016/712).
We thank Giacomo Moretti, Yipin Su and Giuseppe Zurlo for most helpful inputs.



\end{document}